\documentclass[11pt]{article}
\usepackage{latexsym}
\usepackage{amsfonts}
\usepackage{amsmath}
\usepackage[dvips]{graphicx}

\newtheorem{theorem}{Theorem}
\newtheorem{lemma}[theorem]{Lemma}

\newtheorem{claim}[theorem]{Claim}
\newtheorem{obs}[theorem]{Observation}
\newenvironment{proof}{\noindent{\bf Proof.} \enspace }{\noindent $\Box$}

\newcommand{\ints}{{\mathbb Z}}
\newcommand{\rnz}[1]{\mathchoice{\mbox{\rm rnz}}{\mbox{\rm rnz}}%
    {\mbox{\the\scriptfont0 rnz}}{\mbox{\the\scriptfont0 rnz}}(#1)}
\newcommand{\lcabt}{\mbox{\rm LCA}_{BT}}

\begin{document}

\title{\LARGE The Euler Path to Static Level-Ancestors}

\author{
Amir M.~Ben-Amram\\ {\tt amirben@mta.ac.il}
}

\maketitle

\begin{abstract}

Suppose that a rooted tree $T$ is given for preprocessing.
The {\em level-ancestor problem\/} is to
answer quickly queries of the following form. Given a vertex
$v$ and an integer $i > 0$, find the $i$th vertex on the path from
the root to $v$.
Algorithms that achieve a
linear time bound for preprocessing and a
constant time bound for a query have been published by
Dietz (1991), Alstrup and Holm (2000), and Bender and Farach (2002).
The first two algorithms address dynamic versions of the problem; the last
addresses the static version only and is the simplest so far.
The purpose of this note is to expose another simple algorithm, derived
from a complicated PRAM algorithm by Berkman and Vishkin (1990,1994).
We further show some easy extensions of its functionality,
adding queries for descendants and \emph{level successors} 
as well as ancestors,
extensions for which the formerly known algorithms are less suitable.
\end{abstract}


\section{Introduction}

The {\em level-ancestor} problem is defined as follows.
Suppose that a rooted tree $T$ is given for preprocessing.
Answer quickly queries of the following form. Given a vertex
$v$ and an integer $i$, find an ancestor
of $v$ in $T$ whose level is $i$,
where the level of the root is 0.

Two related tree queries are: Level Successor---given $v$, find the next vertex
(in preorder) on the same level. Level Descendant---given $v$ and $i$, find the
first descendant of $v$ on level $i$ (if one exists).

The level-ancestor problem is a relative of the better-known LCA (Least Common
Ancestor) problem. In their seminal paper on LCA problems~\cite{HT:84},
Harel and Tarjan solve the level ancestor problem on certain special trees
as a subroutine of an LCA algorithm.
An application of the Level Ancestor problem
is mentioned already in~\cite{Alon-Schieber-87},
although an implementation of this data structure had not yet been published
at the time.

The first published algorithms for the level ancestor
problem were a PRAM algorithm
by Berkman and Vishkin~\cite{BV89, BV94}, and a serial (RAM) algorithm
by Dietz~\cite{Di91} that accommodates dynamic updates.
Alstrup and Holm~\cite{AH00} gave an algorithm that solves an extended dynamic
problem, and has the additional advantage that its static-only version is
simpler than the previous algorithms.
Finally, the simplest algorithm---for the static problem
only---was given by Bender and Farach~\cite{BF04}.

It is curious that very complicated algorithms to
address theoretical challenges, namely dynamization and parallelization,
had been
published for this problem
earlier than any simple algorithm for the most basic and
useful variant (static, on serial RAM). It is also curious that
the essential ideas
for such an algorithm do appear in Berkman and Vishkin's solution but this
potential contribution was missed, since they concentrated on the PRAM
problem, for which they gave a notoriously impractical algorithm
(involving a tableof almost $2^{2^{28}}$ entries).
The first goal of this paper is to rectify this situation by
presenting a sequential algorithm based on the approach of Berkman and
Vishkin. This is not done just for historical interest, but because the
algorithm here presented is simply useful: it is efficient and easy to
implement (and has been implemented).
Furthermore, we shall present
a few useful extensions that were either unsupported by previous
work, or supported in much more complicated ways.
%
Specifically, we show how to
accommodate level successor and level descendant queries, in addition to
level ancestor.
Together, these two queries are useful for iterating over
the descendants of a vertex at a given level.
For example applications of the extension,
see~\cite{YuanAtallah08,YuanAtallah09}.

\paragraph{Technical remarks.}
Since we only consider data structures that support $O(1)$-time queries,
we refer to the algorithms by the preprocessing cost. That is,
an $O(n)$-time algorithm means linear-time preprocessing.
The data to the algorithm is a tree $T$ whose precise representation
is of little consequence (since standard representations are
interchangeable within linear time). We assume that vertices are
identified by numbers 0~through $n-1$.

\section {The Euler Tour and the Find-Smaller problem}
\label{sec-fs}

Like the better-known LCA algorithm that also originates from~\cite{BV89},
this Level Ancestor algorithm is based on the following key ideas:
\begin{itemize}
\item
The {\em Euler Tour\/} representation of a tree reduces the problem 
to a problem on a linear array.
\item
A data structure with $\Theta(n\log n)$ preprocessing time (and size)
is given for this problem.
\item
This solution is improved to linear-time preprocessing and size using the
{\em microset technique\/}~\cite{GT:85, HT:84}.
\end{itemize}
The microset technique is also used in other work on level ancestors~\cite{AH00,BF04,
MunroRao04,GRR04} but they all apply at least part of the processing to the {\em tree},
using various methods of decomposition into subtrees.
Here, all processing is applied to the Euler-tour array.

Consider a tree $T=(V,E)$, rooted at some vertex $r$. 
For each edge $(v \rightarrow u)$ in $T$, add its anti-parallel edge
$(u \rightarrow v )$. This results in a directed graph $H$.
Since the in-degree and out-degree of each vertex of $H$ are the same,
$H$ has an Euler tour that starts and ends in the root $r$ of $T$.
Note that the tour consists of $2(n-1)$ arcs, hence $2n-1$ vertices
including the endpoints.

By a straight-forward application of DFS on $T$ we can compute the
following information:
\begin{enumerate}
\item
An array $E[0..2n-2]$ such that $E[i]$ is the $i$th vertex on the Euler tour.
\item
An array $L[0..2n-2]$ such that $L[i]$ is the level 
of the $i$th vertex on the Euler tour.
\item
An array $R[0..n-1]$ such that $R[v]$ is the index of the last occurrence
of $v$ in the array $E$, called the {\em representative\/} of $v$.
\end{enumerate}

\smallskip\par\noindent
\begin{obs} \label{obs-reduction}
Let $l < \mbox{\rm level}(v)$.
Vertex $u$ is the level-$l$ ancestor of vertex $v$ if and only if 
$u$ is the first vertex after the last occurrence of $v$ in the Euler
tour such that $\mbox{\rm level}( u ) \le l$.
\end{obs}

By this observation, the computation of the arrays $E$, $L$ and $R$
reduces the level-ancestor problem to the following

\paragraph*{FIND-SMALLER (FS) Problem.}\ \par
\smallskip\noindent
{\em Input for preprocessing}:
Array $A = ( a_1 , a_2 , \ldots , a_n ) $ of integers 
\par\smallskip\noindent
{\em Query}:
Let $0 \le i < n$ and $x\in \ints$.
A query $\mbox{\rm FS}_A(i,x)$ seeks
the minimal $j > i$ such that $a_j \leq x$. If no such 
$j$ exists, the answer is 0.

\smallskip\noindent
Our goal is to preprocess the array $A$ so that 
each query $\mbox{\rm FS}_A(i,x)$ can be processed in $O(1)$ time.

The Euler tour implies that the difference between successive elements 
of array $L$ is exactly one. Therefore, for our goal, it suffices
to solve the following restricted problem:
\smallskip\par\noindent
$(\pm 1)$FS is the Find-Smaller problem restricted to arrays $A$ where
for all $i$, $|a_i - a_{i+1}| = 1$.

We remark that the general Find Smaller problem cannot be solved with
$O(1)$ query time, if one requires a polynomial-space data structure,
and assumes a polylogarithmic word length; the reason is that the static
predecessor problem, for which non-constant lower bounds are
known~\cite{BeameFich02}, can be easily reduced to it.

Another preparatory definition is the following. Let $n$ be a power of
two and consider a balanced binary tree of $n-1$ nodes numbered $1$
through $n-1$ in symmetric order (thus, 1 is the leftmost leaf and $n-1$ the
rightmost).
The height of node $i$ is $\rnz i$, the position of the
rightmost non-zero bit in the binary representation of $i$, counting from 0.
 We denote by $\lcabt(i,j)$ the least common ancestor of nodes $i$ and $j$.
For the algorithms,
we assume that $\lcabt(i,j)$ is computed in constant time.
In fact, it can be computed using standard
machine instructions and the MSB (most significant set bit)
function; this function is implemented as an
instruction in many processors, but could also be provided by
a precomputed table.
Following is a useful property of the $\rnz$ function.
\begin{lemma} \label{lem-rnz}
If $j < i$ are two nodes of the complete binary tree, and
$k=\lcabt(j,i)$, then $i-j+1 \le 2^{1+\rnz k}$, and $i-k+1 \le 2^{\rnz k}.$
\end{lemma}
We omit the easy proof. Finally, for uniformity of notation, we define
$\lcabt(j,i)$ for $j\le 0 < i$ to be 0.



\section{Basic constant-time-query algorithm}

In this section we describe an $O(n\log n)$-time preprocessing algorithm for 
the $(\pm 1)$FS problem.
Throughout this section and the sequel,
we make the simplifying assumption that $n$ is a power of two.

Our description of the Basic algorithm has two steps.
(1) The output of the preprocessing algorithm is specified, and
it is shown how to process an FS query in constant time
using this output.
(2) The preprocessing algorithm is described.
This order helps motivating the presentation.

\subsection{Data structure and query processing}

For each $i$, $0 \le i < n$,
the preprocessing algorithm constructs an array $B_i[1..f(i)]$,
where $f(0)=n$ and for $i>0$, $f(i)=3\cdot 2^{{\rnz i}}$.
%
In $B_i[j]$ we store the answer to $\mbox{\rm FS}(i, a_i-j)$.

A query $\mbox{\rm FS}(i,x)$ is processed as follows
(we assume that $x>a_0-n$, for otherwise the answer is immediate,
due to the $\pm 1$ restriction).
\\
(1) If $x\ge a_i$, return $i$.
\\
(2) Let $d = a_i-x$.
    If $d\le f(i)$ return $B_i[d]$. 
\\
(3) Otherwise, let $k = \lcabt(i-d+1, i)$; return $B_k[a_k - x]$.

Figure~\ref{fig-wt} demonstrates the structure for 
a 16-element array $A$,
except that all the arrays $B_i$ are truncated to 8 elements.
In this example, the query $\mbox{\rm FS}(6,3)$ is answered immediately as
$B_{6}[1] = 11$; the query $\mbox{\rm FS}(9,1)$ is answered via Case~(3):
$k=8$ and $B_k[a_k-1] = B_8[5] = 13$.

 
\begin{figure}[t]
$$\begin{array}{ccccccccccccccccc}
i  & 0 & 1 & 2 & 3 & 4 & 5 & 6 & 7 & 8 & 9 & 10 & 11 & 12 & 13 & 14 & 15 \\
a_i  & 0 & 1 & 2 & 3 & 2 & 3 & 4 & 5 & 6 & 5 &  4 &  3 &  2 &  1 &  2 &  1 \\
~  &  ~&  ~&  ~&  ~&  ~&  ~&  ~&  ~&  ~&  ~&  ~ &  ~ &  ~ &  ~ &  ~ &  ~ \\
j & B_0 & B_1 & B_2 & B_3 & B_4 & B_5 & B_6 & B_7 & B_8 & B_9 & B_{10}
 & B_{11} & B_{12} & B_{13} & B_{14} & B_{15}\\[2pt]
1  & 0 & 0 & 13&  4& 13&12& 11& 10&  9& 10&  11&  12& 13 &  0 & 15 &  0 \\
2  & 0 & 0 & 0 & 13&  0&13& 12& 11& 10& 11&  12&  13&  0 &  0 &  0 &  0 \\
3  & 0 & 0 & 0 &  0&  0& 0& 13& 12& 11& 12&  13&   0&  0 &  0 &  0 &  0 \\
4  & 0 & ~ & 0 &  ~&  0& ~&  0&  ~& 12&  ~&   0&   ~&  0 &  ~ &  0 &  ~ \\
5  & 0 & ~ & 0 &  ~&  0& ~&  0&  ~& 13&  ~&   0&   ~&  0 &  ~ &  0 &  ~ \\
6  & 0 & ~ & 0 &  ~&  0& ~&  0&  ~&  0&  ~&   0&   ~&  0 &  ~ &  0 &  ~ \\
7  & 0 & ~ &  ~&  ~&  0& ~&  ~&  ~&  0&  ~&   ~&   ~&  0 &  ~ &  ~ &  ~ \\
8  & 0 & ~ &  ~&  ~&  0& ~&  ~&  ~&  0&  ~&   ~&   ~&  0 &  ~ &  ~ &  ~ \\
\end{array}$$
\caption{The basic FS structure.}
\label{fig-wt}
\end{figure}

We now explain the algorithm. Correctness of Case~(2) is obvious by the
definition of the structure. The correctness in Case~(3) hinges on two
claims.
The first, Claim~\ref{clm-withinrange} below,
shows that the reference to $B_k[a_k-x]$ is within 
bounds; the second, Claim~\ref{clm-backup},
shows that the answer found there is the right one.

\begin{claim} \label{clm-withinrange}
In Case~(3), we have $0 < a_k-x \le f(k)$.
\end{claim}

\begin{proof}
For the first inequality: $k > i-d$ by its definition;
we are dealing with $(\pm 1)$FS, therefore
$a_k > a_i - d = x$.
For the second inequality: 
We assume $k>0$, as for $k=0$ and the claim clearly holds.
Consider the complete binary tree of $n-1$ nodes, used to define
$\lcabt(i,j)$. The algorithm sets $k = \lcabt(i-d+1, i)$, so 
by Lemma~\ref{lem-rnz},
\begin{eqnarray*}
2^{1+ \rnz k} &>& i - (i-d+1) = d-1 = a_i - x -1\\ 
2^{\rnz k}    &>& i-k\\
\Rightarrow\qquad 3\cdot 2^{\rnz k} &\ge& a_i - x + i - k .\\
\end{eqnarray*}
Since the difference between consecutive elements is $\pm 1$, we have
$a_k \le a_i + i-k$, so we conclude that
$$3\cdot 2^{\rnz k} \ge a_k - x.$$
\end{proof}

\begin{claim} \label{clm-backup}
If $i-a_i+x < k\le i$, then $\mbox{\rm FS}(k,x) = \mbox{\rm FS}(i,x)$.
\end{claim}

\begin{proof}
Because we are dealing with $(\pm 1)$FS,
the values $a_k,\dots,a_i$ are all in the interval $(a_i-(i-k), a_i+(i-k))$.
By assumption we have $a_i-(i-k) > x$. Thus, the answer to 
$\mbox{\rm FS}(k,x)$ lies beyond $a_i$, and is also the answer to
$\mbox{\rm FS}(i,x)$.
\end{proof}

\subsection{The preprocessing algorithm}

It is easy to verify that the size of the data structure is $\Theta(n\log n)$.
To construct it in $O(n\log n)$ time, we perform a sweep from right to left;
that is, for $i=n-1,n-2,\dots,0$
we compute an array $F[\min A,\dots,\max A]$
where $F[x]$ is the index of the first $j>i$ such that $a_j=x$ (or 0
by default). Note that this is not the same as
$\mbox{\rm FS}(i,x)$.
Initializing $F$ for $i=n-1$ is trivial and that
updating it when $i$ is decremented is constant-time. For each $i$, $B_i$
is just a copy of an appropriate section of $F$. This completes the
preprocessing.

\section{Improved constant-time-query algorithm}

In this section we describe an $O(n)$-time algorithm, based on the solution
of the former section together with the {\em microset technique}. The
essence of the technique is to fix a {\em block length} 
$b=\lfloor (\log n)/2\rfloor$ 
and to sparsify the structure
of the last section by using it only on block boundaries, reducing its
cost to $O(n)$, while for {\em intra-block queries\/} we use an additional
data structure, the {\em micro-structure}. For presentation's sake,
we now provide a specification of the micro-structure and go on to describe the 
rest of the structure. The implementation of the micro-structure will be
dealt with in the following section. 

For working with blocks, without resorting to numerous division operators,
we shall write down some numbers (specifically, array indices)
in a quotient-and-remainder notation,
$ib+j$, where it is tacitly assumed that $0\le j <b$.

\paragraph*{The Micro-Structure.} This data structure is assumed to support
in $O(1)$ time the following query: $\mbox{\rm Micro.FS}(ib+j,x)$---return
the answer to $\mbox{\rm FS}(ib+j,x)$ provided that it is less than $(i+1)b$.
Otherwise, return 0.

\paragraph*{The FS Structure.} 
For each $i$, $0 \le i < n/b$,
our preprocessing algorithm now constructs two arrays:
\begin{enumerate}
\item
A {\em near\/} array $N_i[1,\dots,2b]$ 
such that $N_i[j]$ stores 
the answer to $\mbox{\rm FS}(ib, a_{ib}-j)$
 (namely, the first $2b$ entries of $B_{ib}$ of the previous section).
\item
A {\em far\/} array $F_i[1,\dots,f(i)]$ such that
$$F_i[j] = \lfloor \mbox{\rm FS}(ib, a_{ib} - jb) / b\rfloor$$
\end{enumerate}
Thus, the arrays are not only sparsified, but also (for the far arrays)
are their values truncated.
Referring to the example in Figure~\ref{fig-wt}, we have $b=2$,
so near arrays have 4 elements, e.g., $N_{4}=(9,10,11,12)$.
The far array $F_{4}$ has $f(4)=12$ entries:
$(5,6,0,\dots,0)$.

The following fact follows from the $(\pm 1)$ restriction and the
definition of $F_i$:
\begin{obs} \label{obs-F}
If $F_i[j] = k$, then 
$ a_{ib} - jb \le a_{kb} <  a_{ib} - (j-1)b$.
\end{obs}

\paragraph*{Query processing.}
A query $\mbox{\rm FS}(ib+j,x)$ is processed as follows
(we assume once again that $x>a_0-n$).
\renewcommand{\labelenumiii}{{(\arabic{enumi}.\arabic{enumii}.\arabic{enumiii})}}
\renewcommand{\theenumii}{{.\arabic{enumii}}}
\renewcommand{\theenumiii}{{.\arabic{enumiii}}}
\renewcommand{\labelenumii}{{(\arabic{enumi}.\arabic{enumii})}}
\renewcommand{\labelenumi}{{(\arabic{enumi})}}
\begin{enumerate}
\item
  If $x\ge a_{ib+j}$, return $ib+j$.
\item \label{round}
  If $j=0$ then
  \begin{enumerate}
  \item \label{close}
     If $x\ge a_{ib}-2b$ return $N_i[a_{ib}-x]$
  \item
    $d \gets {\lfloor (a_{ib}- x)/b\rfloor}$; 
    
    if $d\le f(i)$ then
    \begin{enumerate}
    \item \label{simple}
      $k\gets F_i[d]$; return $\mbox{\rm FS}(kb,x)$.
      
   \makebox[0pt][r]{else\ }
   \item \label{notsimple}
  $l\gets i-d+1$; $k\gets \lcabt(l,i)$; return $\mbox{\rm FS}(kb,x)$.
   \end{enumerate}
   \end{enumerate}
\item \label{notround}
  (if $j>0$)\\
  $m\gets \mbox{\rm Micro.FS}(ib+j,x)$;\\
  if $m\ne 0$, return $m$, else return $\mbox{\rm FS}((i+1)b,x)$.
\end{enumerate}

\noindent
The following observations
justify this procedure,  and also show that there is no real
recursion here: the recursive calls can
actually be implemented as {\tt goto}s and they never loop.
\begin{enumerate}
\item
In Case~(\ref{notround}), when the micro-structure does not yield the
answer, it follows that the element sought is further than
$(i+1)b$; therefore the recursive call is correct, and will be handled at
Case~(\ref{round}).
\item
In Case~(2.2.1), we have (see Observation~\ref{obs-F} above)
$$a_{kb} < a_{ib} - (d-1)b < x + 2b$$
and
$$a_{kb} \ge a_{ib} - db \ge x$$
Therefore, the recursive call is handled correctly at Case~(\ref{close}).
\item
For Case~(2.2.2), we can show, as for the basic algorithm,
that $\mbox{\rm FS}(kb,x) = \mbox{\rm FS}(ib,x)$ (same proof as before),
and that $0 < a_{kb}-x \le f(k)\cdot b$, showing that
the recursive call falls back to Case~(2.2.1).
The last inequality is proved 
as Claim~\ref{clm-withinrange2}.
\end{enumerate}

\begin{claim} \label{clm-withinrange2}
In Case~(2.2.2), we have $a_{kb}-x \le f(k)\cdot b$.
\end{claim}

\begin{proof}
We assume $k>0$.
By Lemma~\ref{lem-rnz},
\belowdisplayskip=0pt plus 3pt minus 0pt
\belowdisplayshortskip=0pt plus 3pt minus 0pt
\begin{eqnarray*}
2^{1+ \rnz k} &\ge& i - l + 1 = d > \frac{a_{ib} - x}{b} - 1 \\
2^{\rnz k}    &\ge& i-k+1\\
\Rightarrow\quad f(k) = 3\cdot 2^{\rnz k} &>& \frac{a_{ib} - x}{b} + i - k
              = \frac{a_{ib} - x + ib - kb}{b} 
            \ge \frac{a_{kb} - x}{b},\\
\end{eqnarray*}
where the last inequality is justified by the $(\pm 1)$ property.
\end{proof}

\section{The Micro Structure}

The purpose of the micro structure is to support ``close'' queries,
i.e., 
return the answer to $\mbox{\rm FS}(ib+j,x)$ provided that it is at most 
$(i+1)b$. There are several ways to implement this structure, with subtle
differences in performance or ease of implementation. We describe two.

\subsection{Berkman and Vishkin's structure}
\label{sec-bvmicro}

The basis for fast solution of in-block queries in~\cite{BV94} is observing that,
up to normalization, there are less than $2^b$ different possible blocks.
Normalization amounts to subtracting the first element of the block from
all elements; i.e., moving the ``origin'' to zero. Clearly, a query on
any array $A$, $\mbox{\rm
FS}_A(j,x)$, is equivalent to $\mbox{\rm FS}_{A'}(j,x-a_0)$ where $A'$ is the
normalized form of $A$. The bound $2^b$ 
follows from the $(\pm 1)$
restriction. This also allows us to conveniently represent a
block as a binary string of length $b-1$ (which fits in a word).
We obtain the following solution.
{\par\noindent\em Preprocessing\/}:
For every possible ``small'' array $S$ of size $b$,
beginning with 0, and satisfying
the $(\pm 1)$ restriction, build a matrix $M_S[b\times 2b]$ such that
$M_S[j,x]$ is the answer to $\mbox{\rm FS}_{S}(j,x)$ for every $0\le j<b$
and $-b < x < b$. As an identifier of $S$ (to index the array of matrices)
we use the $(b-1)$-bit representation of $S$. While preprocessing an array $A$
of size $n$ for FS queries, we store for every $0\le i < n/b$ the
identifier $S[i]$ of the block $(a_{ib},\dots, a_{(i+1)b-1})$.
{\par\noindent\em Query\/}:
$\mbox{\rm Micro.FS}_{A}(ib+j,x)$ is answered by looking up 
$M_{S[i]}[j, x-a_{ib}]$ (returning 0 if the
second index is out of range).
{\par\noindent\em Complexity\/}:
The query is obviously constant-time. For the preprocessing, creating the
idntifier array $S$ clearly takes $\Theta(n)$ time. The construction of a
single matrix $M_S$ can be done quite simply in $\Theta(b^2)$ time, and
altogether we get $2^b\cdot\Theta(b^2) = O(n)$ time and space.

\subsection{A solution after Alstrup, Gavoille, Kaplan and Rauhe}
\label{sec:alstrup}

Another implementation of the micro structure is suggested by an idea
from~\cite{Alstrup-et-al:04}. In its basic form, as we next describe,
it is really independent of the division into blocks---except that it only
supports queries where the answer is close enough to the query index.

For $i < j\le n$, let
$$m(i,j) = \left\{\begin{array}{ll}
1 & \mbox{if } a_j < \min\{a_i,\dots,a_{j-1}\} \\
0 & \mbox{\rm otherwise.}
\end{array}\right.$$
From the $(\pm 1)$ property, one can easily deduce that 
$\mbox{\rm FS}_{A}(i,a_i-k)$ is precisely the position of the $k$th 1 in
the sequence
$$m(i,i+1), m(i,i+2), \dots$$
The solution to the micro-structure problem, based on this observation,
follows:
{\par\noindent\em Preprocessing\/}:
For every $0\le i < n$, compute and store in an array entry $M[i]$ the
$b$-bit mask $(m(i,i+1),\dots,m(i,i+b))$.

{\noindent\em Query\/}:
$\mbox{\rm Micro.FS}_{A}(i,x)$ is answered (for $x<a_i$)
by looking up the $(a_i-x)$'th set bit in $M[i]$.
The answer is 0 if there is no such bit.

This query returns answers in positions up to $i+b$, rather than
$\lceil i/b\rceil\cdot b$, which can possibly result in a faster query.
As an additional advantage, $b$ can be enlarged up to the word size,
saving both time and space
(there is a certain caveat---see below).

{\noindent\em Query Complexity\/}:
The query is constant-time if we have a constant-time implementation of
the function that locates the $i$'th bit set in a word. In the absence of
hardware support, a precomputed table, of size $O(2^b\cdot b)$, can be used
(but this requires limiting the value of $b$ as before)%
\footnote{Another way,
which is not constant-time in the RAM model,
is to search for this bit using available
arithmetic/logical instructions. Since this is a tight loop
without any memory access, it may be even faster than a table access
on a real computer.}%
.
{\par\noindent\em Preprocessing Algorithm\/}:
To compute the mask array $M$, we scan $A$ from right to left while
maintaining two pieces of data: the mask corresponding to the current
position $i$, and a stack that includes the indices $\{j\mid m(i,j)=1\}$
up to the end of $A$. Each time the current position $i$ is decremented,
$i$ is pushed unto the
stack, possibly kicking off the top two elements (specifically,
if $a_{i+1}=a_i+1$). The current mask is easily adjusted in $O(1)$ time.

Clearly, the computation of $M$ takes $\Theta(n)$ time, and this is also the space
required. Fischer and Heun~\cite{FH:escape07} propose to
apply this technique within microblocks;
in other words,
revert to the Berkman-Vishkin approach of maintaining a table indexed by
the block identifier, but keep the mask table instead of an explicit
answer matrix. This saves a factor of $b$ in the size of the micro structure,
but is likely to
be competitive in speed only if the bit-finding operation we make
use of is supported by hardware.

\subsection{Saving memory}

In our description of the algorithm we aimed for simplicity while achieving
the desired asymptotic bounds: constant-time query together with $O(n)$ space
and preprocessing time. If, for some practical reason, the constant in the 
$O(n)$ space bound is of importance, one can look for improvements, which
are not hard to find. We list two simple constant-factor improvements.
\begin{enumerate}
\item
The size $f(i)$ of $B_i$ can be defined to be
$2\cdot 2^{\rnz i} + 1$ instead of $3\cdot 2^{\rnz i}$.
Moreover, assuming that all $a_i\ge 0$ (as is the
case when using FS to solve Level Ancestors), we can use
$\min(f(i),a_i)$. This eliminates $B_0$, and may give additional savings
further on,
depending on the shape of the tree in the Level Ancestor problem.
\item
The size of the arrays $E$, $L$ in the reduction of Level Ancestors to
the Find-Smaller problem can be cut in half by listing a vertex $v$
in $E$ only when visited by the Euler Tour for the last time
(put otherwise, we list the vertices in post-order).
It is still true that the level-$l$ ancestor of $v$
is the first vertex $u$ occurring after $v$
such that $\mbox{\rm level}( u ) \le l$. Thus, the reduction to Find
Smaller is still correct. However,
now the FS problem that results does not enjoy the
$(\pm 1)$ property. But it has a similar property: for all $i$,
$a_{i+1}\ge a_i+1$. Interestingly, this suffices for implementing the
algorithm, at least with the micro-structure of Section~\ref{sec:alstrup}.
Thus, this saving in memory incurs no loss in running time.
\end{enumerate}

\paragraph*{Remark.}
This part of the solution is 
where the simplification with respect to~\cite{BV94} is
most significant, although the outline (initial, non-optimal, solution,
and usage of micro-blocks) is similar.

\section{The Level-Descendant and Level-Successor Queries}
\label{sec-ld}

Observation~\ref{obs-reduction} can easily be turned from 
ancestors to descendants:
\smallskip\par\noindent
\begin{obs} 
Let $l > \mbox{\rm level}(v)$.
Vertex $u$ is the first level-$l$ descendant of vertex $v$ if and only if 
$u$ is the first vertex after the first occurrence of $v$ in the Euler
tour such that $\mbox{\rm level}( u ) \ge l$, provided that this vertex is
a descendant of $v$. If it is not, $v$ has no level-$l$ descendant.
\end{obs}

By this observation, the level descendant query
reduces to a Find-Greater problem, analogous to Find-Smaller and solved in
the same way, plus a test of descendance.
Thus, to add this functionality, we use the same arrays $E$, $L$ and add a
vector $F$ maintaining the first occurrence of each vertex in the tour.
We also
need a search structure for ``Find Greater.'' This structure is, of course, completely
symmetric to the Find-Smaller structure so no further explanation should be
necessary (incidentally, the micro
table {\`a-la} Berkman-Vishkin can be shared).
Testing for descendance is easy---$u$ descends from $v$ if and only if
$F[v]\le F[u]\le R[v]$.

The level successor query is handled similarly, by the following observation:
\begin{obs} 
Vertex $u$ is the level successor of vertex $v$ if and only if 
$u$ is the first vertex after the last occurrence of $v$ in the Euler
tour such that $\mbox{\rm level}( u ) \ge \mbox{\rm level}( v )$.
\end{obs}

\section{Conclusion} 

I described how to construct and query a data structure for answering
Level Ancestor queries on trees. The algorithm is based on Berkman and Vishkin's Euler Tour technique
and is, in essence, a simplification of their PRAM algorithm.
In contrast to the original, this version of the algorithm 
is simple and practical. The algorithm was implemented
in C by Victor Buchnik;
the code can be obtained from Amir Ben-Amram.

Another advantage of this algorithm is that it can be
easily extended to support queries for Level
Descendants and Level Successors.

\bibliographystyle{abbrv}
\bibliography{amir}

\begin{thebibliography}{10}

\bibitem{Alon-Schieber-87}
N.~Alon and B.~Schieber.
\newblock Optimal preprocessing for answering on-line product queries.
\newblock Technical report, Tel Aviv University, 1987.

\bibitem{Alstrup-et-al:04}
S.~Alstrup, C.~Gavoille, H.~Kaplan, and T.~Rauhe.
\newblock Nearest common ancestors: {A} survey and a new algorithm for a
  distributed environment.
\newblock {\em Theory of Computing Systems}, 37(3):441--456, 2004.

\bibitem{AH00}
S.~Alstrup and J.~Holm.
\newblock Improved algorithms for finding level ancestors in dynamic trees.
\newblock In U.~Montanari, J.~D.~P. Rolim, and E.~Welzl, editors, {\em
  Proceedings of the 27th International Colloquium on Automata, Languages and
  Programming (ICALP)}, volume 1853 of {\em LNCS}, pages 73--84.
  Springer-Verlag, July 2000.

\bibitem{BeameFich02}
P.~Beame and F.~E. Fich.
\newblock Optimal bounds for the predecessor problem and related problems.
\newblock {\em J. Comput. Syst. Sci}, 65(1):38--72, 2002.

\bibitem{BF04}
M.~A. Bender and M.~Farach-Colton.
\newblock The level ancestor problem simplified.
\newblock {\em Theor. Comput. Sci}, 321(1):5--12, 2004.

\bibitem{BV89}
O.~Berkman and U.~Vishkin.
\newblock Recursive *-tree parallel data-structure.
\newblock In {\em 30th Annual Symposium on Foundations of Computer Science},
  pages 196--202. {IEEE}, 1989.

\bibitem{BV94}
O.~Berkman and U.~Vishkin.
\newblock Finding level-ancestors in trees.
\newblock {\em J. Computer and System Sciences}, 48(2):214--230, 1994.

\bibitem{Di91}
P.~F. Dietz.
\newblock Finding level-ancestors in dynamic trees.
\newblock In {\em Workshop on Algorithms and Data Structures (WADS)}, pages
  32--40, 1991.

\bibitem{FH:escape07}
J.~Fischer and V.~Heun.
\newblock A new succinct representation of rmq-information and improvements in
  the enhanced suffix array.
\newblock In {\em Proceedings of the International Symposium on Combinatorics,
  Algorithms, Probabilistic and Experimental Methodologies ({ESCAPE}'07)},
  volume~?? of {\em Lecture Notes in Computer Science}, pages ??--?? Springer,
  2007.

\bibitem{GT:85}
{Gabow, Harold N.} and {Tarjan, Robert E.}
\newblock A linear time algorithm for a special case of disjoint set union.
\newblock {\em J. Comput. Syst. Sci.}, 30:209--221, 1985.

\bibitem{GRR04}
R.~F. Geary, R.~Raman, and V.~Raman.
\newblock Succinct ordinal trees with level-ancestor queries.
\newblock In J.~I. Munro, editor, {\em Proceedings of the Fifteenth Annual
  {ACM}-{SIAM} Symposium on Discrete Algorithms, {SODA} 2004, New Orleans,
  Louisiana, {USA}, January 11-14, 2004}, pages 1--10. SIAM, 2004.

\bibitem{HT:84}
D.~Harel and R.~E. Tarjan.
\newblock Fast algorithms for finding nearest common ancestors.
\newblock {\em SIAM Journal on Computing}, 13(2):338--355, May 1984.

\bibitem{MunroRao04}
J.~I. Munro and S.~S. Rao.
\newblock Succinct representations of functions.
\newblock In J.~D{\'i}az, J.~Karhum{\"a}ki, A.~Lepist{\"o}, and D.~Sannella,
  editors, {\em Automata, Languages and Programming: 31st International
  Colloquium, {ICALP} 2004, Turku, Finland, July 12-16, 2004. Proceedings},
  volume 3142 of {\em Lecture Notes in Computer Science}, pages 1006--1015.
  Springer, 2004.

\bibitem{YuanAtallah08}
H.~Yuan and M.~J. Atallah.
\newblock Efficient distributed third-party data authentication for tree
  hierarchies.
\newblock In {\em 28th IEEE International Conference on Distributed Computing
  Systems (ICDCS '08)}, pages 184--193. IEEE Computer Society, 2008.

\bibitem{YuanAtallah09}
H.~Yuan and M.~J. Atallah.
\newblock Efficient data structures for range-aggregate queries on trees.
\newblock In R.~Fagin, editor, {\em Database Theory---{ICDT} 2009, 12th
  International Conference, St. Petersburg, Russia, Proceedings}, volume 361 of
  {\em ACM International Conference Proceeding Series}, pages 111--120. ACM,
  2009.

\end{thebibliography}

\end{document}